\documentclass[twocolumn,showpacs,preprintnumbers,amsmath,amssymb]{revtex4}

\usepackage{graphicx}
\usepackage{dcolumn}
\usepackage{bm}
\usepackage{graphicx}
\usepackage{epsfig}
\include{graphic}

\begin{document}
\preprint{APS/123-QED}
\title{Resonances in Ferromagnetic Gratings Detected by Microwave Photoconductivity}

\author{Y. S. Gui, S. Holland, N. Mecking, and C. -M. Hu\footnote{Electronic address: hu@physnet.uni-hamburg.de}}

\affiliation{Institut f\"ur Angewandte Physik und Zentrum f\"ur
Mikrostrukturforschung, Universit\"at Hamburg, Jungiusstra\ss e
11, 20355 Hamburg, Germany}

\date{\today}

\begin{abstract}

We investigate the impact of microwave excited spin excitations on
the DC charge transport in a ferromagnetic (FM) grating. We
observe both resonant and nonresonant microwave photoresistance.
Resonant features are identified as the ferromagnetic resonance
(FMR) and ferromagnetic antiresonance (FMAR). A macroscopic model
based on Maxwell and Landau-Lifschitz equations reveals the
macroscopic nature of the FMAR. The experimental approach and
results provide new insight in the interplay between photonic,
spintronic, and charge effects in FM microstructures.

\end{abstract}

\pacs{73.50.Pz, 41.20.Jb, 76.50.+g, 42.79.Dj}

%
%
%
%
%
%

\maketitle The connection between the DC and high frequency
response of the metal to external fields looks like a one-way
path. On the one hand, it is text book knowledge that due to the
ohmic dissipation, the DC conductivity $\sigma_{0}$ determines the
skin depth $\delta=\sqrt{2/\mu_{0}\sigma_{0}\omega}$ of the
electromagnetic radiation with the frequency $\omega = 2\pi f$,
where $\mu_0$ is the permeability of vacuum. On the other hand,
there is little knowledge about the inverse effect of the high
frequency response on the DC transport in metals, which is in
contrast to the case of semiconductors, where a whole zoo of
photoconductivity phenomena, ranging from the intrinsic,
extrinsic, to the bolometric effect, are all based on such an
influence.

Recently, a breakthrough has been achieved in ferromagnetic (FM)
metals. By combining the giant magnetoresistance effect of a FM
multilayer with the microwave absorption, high frequency
resonances were detected by measuring the DC resistance
\cite{Tsoi2000}. It was the first photoconductivity experiment on
FM multilayers, which bridged static and dynamic properties, and
paved the way for recent highlights of generating microwave
oscillations by a spin-polarized DC current \cite{Kiselev2003}.
Despite of broad interest in studying the interplay of static and
dynamic responses in FM multilayers, the basic question of the
impact of the high frequency response on the DC transport in a
single layer FM metal remains open.

In this paper, we answer this question by performing microwave
photoconductivity measurements directly on a single layer FM
microstrip. Our primary aim is to explore the bolometric effect
\cite{Hu2003} in the FM metal, which may bridge the high frequency
absorbance $A(\omega)$ with the DC resistance change $\Delta R$
via a simple relation
\begin{equation}
\Delta R = S\cdot A(\omega),
\end{equation}
where $S = \frac{\partial R}{\partial
T}\frac{P_{0}\tau_{e}}{C_{e}}$ is a sensitivity parameter that
depends on the specific heat $C_{e}$ of electrons, the incident
power $P_{0}$ of the radiation, and the energy relaxation time
$\tau_{e}$ of photo-excited spin/charges. The relation was
previously only known for semiconductors \cite{Neppl1979}. We
demonstrate that based on the interplay of the spin dynamics and
the DC charge transport, both the ferromagnetic resonance (FMR)
\cite{Kittel} and ferromagnetic antiresonance (FMAR)
\cite{Yager1949} can be detected by the photoconductivity
technique. Using a model based on Maxwell and Landau-Lifschitz
equations, we reveal the unique macroscopic nature of the FMAR in
the FM grating, which has the potential to integrate spintronic
and photonic features in FM microstructures.

Our experiments are performed on an array of Ni$_{80}$Fe$_{20}$
(Permalloy, Py) microstrip with a width $W$= 50 $\mu $m and a
thickness $d$ = 60 nm. As illustrated in insets of Fig. 1, the
strip has a total length $L \approx$ 10 cm and runs meandering in
a square of about 3$\times$3 mm$^{2}$, forming 30 periods of FM
grating with a period $a$ = 70 $\mu$m. The Py strip is deposited
on a semi-insulating GaAs substrate using photolithography and
lift-off techniques. The DC conductivity $\sigma_{0}$ of the Py
strip is determined to be 3.2 (5.0) $\times$ 10$^{4}$
$\Omega^{-1}$cm$^{-1}$ at 300 (4.2) K. A swept-signal generator is
connected with a circular oversized waveguide, which brings the
microwave radiations with $f$ between 17.5 - 20 GHz down to the
sample set in a cryostat.

\begin{figure} [t]
\begin{center}
\epsfig{file=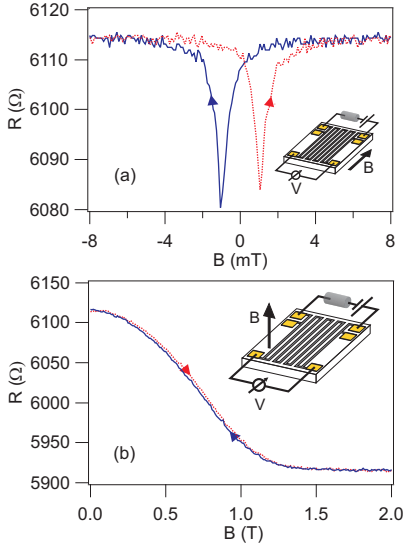,height=7.2 cm} \caption{(color online). (a)
Parallel (\textit{$\theta $} = 0$^{o}$) and (b) perpendicular
(\textit{$\theta $} = 90$^{o}$) AMR effect measured with the
applied magnetic field and current bias as shown in the
inset.}\label{Fig.1}
\end{center}
\end{figure}

Before discussing the photoconductivity of the Py strip, we show
in Fig. 1 the static property of our sample without microwave
radiations. By applying the external magnetic field $ H =
B/\mu_{0}$ along the easy axis parallel (\textit{$\theta $} =
0$^{o}$) to the current flow in the strip, we measure the
anisotropic magnetoresistance (AMR) and plot it \cite{AMR} in Fig.
1(a). The sharp minimum at $\pm$ 1.2 mT corresponds to the
coercive field of the strip \cite{Adeyeye1996}, which increases
with increasing the angle \textit{$\theta $} (not shown). At
\textit{$\theta $} = 90$^{o}$ when the applied B field is along
the hard axis perpendicular to the strip plane, perpendicular AMR
is measured and plotted in Fig. 1(b). The estimated saturation
magnetization ($M_0$) is about 1.2 T/$\mu_{0}$ and the normalized
AMR is about 3.2{\%}, both in agreement with earlier reports
\cite{Adeyeye1996}.

We perform the photoconductivity experiment at \textit{$\theta $}
= 90$^{o}$ in the Faraday configuration with the microwave wave
vector $\textbf{k}\parallel\textbf{B}$. Fig. 2 shows typical
photoresistance traces measured as a function of the B field at
4.2 K for different microwave frequencies. The curves are
vertically offset for clarity. A DC current of $I$ = 90 $\mu $A is
applied. The radiation-induced voltage change $\bigtriangleup V$
is measured via lock-in technique by modulating the microwave
power with a frequency of 123 Hz. The photoresistance
$\bigtriangleup R =\bigtriangleup V/I$ measures the
microwave-induced DC magnetoresistance change of the Py strip. In
addition to a nonresonant background photoresistance at the order
of 10 m$\Omega$, which is about a few ppm of the DC
magnetoresistance $R$ of the Py strip, we observe clearly two
resonances. One appears as a peak and the other as a dip. The
resonant field for both shifts with $f$. We find that
$\bigtriangleup R$ increases with increasing power. The data shown
in Fig. 2 are measured by setting the output power of the
swept-signal generator at 24 dbm, however, the power that reaches
the sample via the long waveguide is significantly reduced. At
17.75 GHz when $f$ approaches the cut off frequency of the
waveguide, $\bigtriangleup R$ is obviously reduced.

\begin{figure} [t]
\begin{center}
\epsfig{file=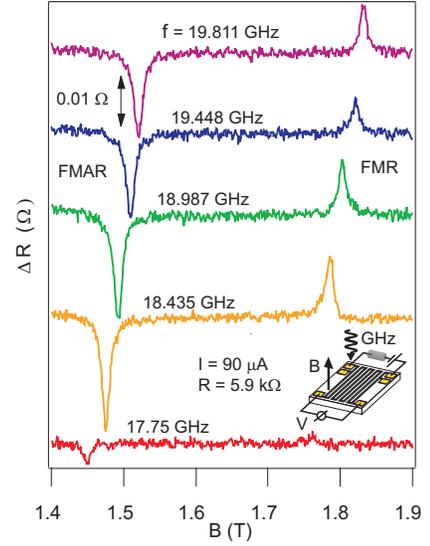,height=7.2 cm} \caption{(color online).
Microwave photoresistance (vertically offset for clarity) of the
Py strip measured as a function of the magnetic field at 4.2 K and
at different microwave frequencies. The inset shows the
measurement configuration.} \label{Fig.2}
\end{center}
\end{figure}

To shed light onto the observed photoconductivity effect, we begin
by analyzing the magnetodynamic response function of our sample.
The dynamic susceptibility tensor $\widehat{\chi}$, which links
the dynamic magnetization $\textbf{m}$ and the dynamic magnetic
field $\textbf{h}$ via $\textbf{m}=\widehat{\chi}\cdot
\textbf{h}$, can be obtained by solving Landau-Lifschitz equation
\cite{Hubert}. For simplicity, we restrict our analysis to the
field range of $H
> M_{0}$ where resonances are observed. Since in our sample $d \ll L, W$,
we start by treating it as a 2D film, taking into account the
demagnetization field but neglecting the anisotropy and the
exchange field.  We get the dynamic permeability tensor
$\widehat{\mu}=\widehat{1}+\widehat{\chi} = \begin{pmatrix}
\mu_{L} & \mu_{T} & 0 \cr -\mu_{T}& \mu_{L} & 0 \cr 0 & 0 & 1
\end{pmatrix}$ with the longditudinal ($\mu_{L}$) and transversal
($\mu_{T}$) complex permeability given by
\begin{eqnarray}
\mu_{L}=1+\frac{\omega_{M}(\omega_{r}-i\alpha\omega)}{(\omega_{r}-i\alpha\omega)^{2}-\omega^{2}},
\cr
\mu_{T}=\frac{i\omega_{M}\omega}{(\omega_{r}-i\alpha\omega)^{2}-\omega^{2}}.
\end{eqnarray}
Here, $\alpha$ is the dimensionless Gilbert damping parameter. We
define $\omega _M = \gamma M_0 $ and $\omega _r = \gamma (H - M_0
)$, with $\gamma = {g\mu _B \mu_{0}}/ \hbar$ the gyromagnetic
ratio which depends on the $g$ factor and the Bohr magneton $\mu
_B$.

The dynamic permeability tensor $\widehat{\mu}$ describes the
gyrotropic response of the FM metal. In the Faraday configuration,
its eigenvalues can be found by solving the equation
$\textbf{k}(\textbf{k}\cdot
\textbf{h})+(k_{0}^{2}\epsilon\widehat{\mu}-\textbf{k}^{2})\textbf{h}
= 0$ deduced from Maxwell equations \cite{Jackson}. We obtain
\begin{equation}
\mu_{\pm}=\mu_{L}\mp i\mu_{T}
=\frac{\omega_{r}+\omega_{M}\mp\omega-i\alpha\omega}{\omega_{r}\mp\omega-i\alpha\omega}
\end{equation}
which define two circular polarized electromagnetic eigenmodes
propagating in the FM film, whose wave vectors are given by
$k_{\pm}^{2}=\epsilon\mu_{\pm}k_{0}^{2}$. Here $\epsilon \approx
i\sigma_{0}/\epsilon_{0}\omega$ is the complex permittivity of the
FM film, $\epsilon_0$, $c$, and $k_{0} = \omega/c$ are the
permittivity, the velocity and the wave vector of light in vacuum.
The $k_{+}$ mode results from the strong coupling of the right
circular electromagnetic wave with the magnetization, which
excites the FMR at the resonant frequency $\omega _r$. The FMR is
inactive for the left circular electromagnetic wave and hence the
$k_{-}$ mode is only weakly influenced by the magnetization.

In Fig. 3(a), we plot the magnetic-field dispersion of the peak
(solid square) and dip (open circle) measured from
photoconductivity spectra. By fitting the dispersion of the peak
using the relation $\omega _r = \gamma (H - M_0 )$, we obtain
$\gamma$ = 183$\mu_{0}$ GHz/T (corresponding to $g$ = 2.08) which
agrees well with the published values \cite{Nibarger2003}, and
$M_{0}$ = 1.15 T/$\mu_{0}$ which is consistent with the value (1.2
T/$\mu_{0}$) estimated from the AMR effect. Therefore we identify
the resonant peak of the photoresistance as the FMR, which has the
microscopic origin of Larmor precession of spins in the FM metal
\cite{Kittel}.

\begin{figure} [t]
\begin{center}
\epsfig{file=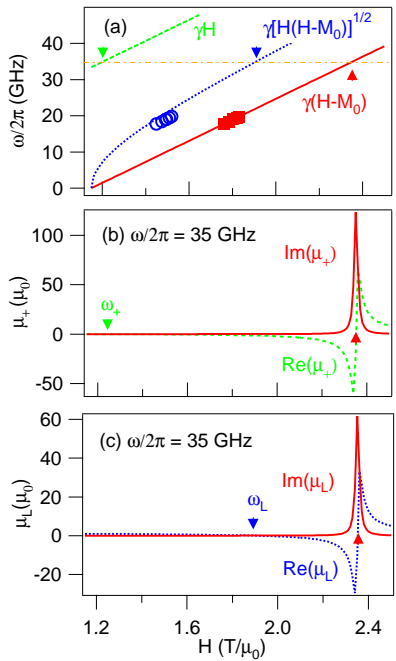,height=9 cm} \caption{(color online). (a)
The measured resonance positions for the photoresistance peak
(solid square) are fitted to the FMR dispersion (solid line). The
measured (open circle) FMAR dispersion is compared with that
calculated using $\mu_{+}$ (dashed line) and $\mu_{L}$ (dotted
curve). (b) $\mu_{+}$ and (c) $\mu_{L}$ are calculated at
$\omega/2\pi$ = 35 GHz, using the parameters $M_0$ = 1.15
T/$\mu_{0}$, $\alpha$ = 0.0075, and $\gamma$ = 183$\mu_{0}$ GHz/T.
Arrows indicate the condition for $Re(\mu)$ = 0.} \label{Fig.3}
\end{center}
\end{figure}

With fitted values for $\gamma$ and $M_{0}$, we calculate and plot
in Fig. 3(b) the B-field dependence of $\mu_{+}$ for $\omega/2\pi$
= 35 GHz. From a line shape fit that we will describe later, we
take $\alpha$ = 0.0075. The real part of $\mu_{+}$ has two zeros.
At the zero located at $\omega = \omega_{r} = \gamma (H - M_0 )$,
which is indicated by the upward arrow in Fig. 3(b), $Im(\mu_{+})$
shows a pole. This is the macroscopic definition of the FMR based
on the magnetodynamic response function. It corresponds to the
condition of resonantly enhanced absorption due to the FMR. For $H
> 0$, $\mu_{-}$ has neither pole nor zero (not shown), because the
FMR is inactive to the left circular polarized electromagnetic
wave.

Note that there is a second zero for $Re(\mu_{+})$ located at
$\omega = \omega_{+} = \gamma H$, which is indicated by the
downward arrow in Fig. 3(b). At this condition, $Im(\mu_{+})$ is
also nearly zero, hence the dynamic susceptibility $\chi_{+}
\simeq$ -1. This is the resonant condition for the FMAR of the FM
film at which the ohmic dissipation due to eddy currents is
suppressed. Early microwave transmission experiments performed on
thick ($d
> \delta$) FM films have confirmed enhanced transmission and reduced
absorption at the FMAR \cite{Yager1949}. One would therefore
attribute the resonant photoresistance dip in Fig. 2 to the FMAR.
However, as shown in Fig. 3(a), the measured resonances for the
dips (open circle) lie far away from the dashed line plotted for
the relation $\omega_{+} = \gamma H$.

The significant discrepancy reflects an intriguing macroscopic
nature of the FMAR in the microstructured FM layer. As shown in
insets of Figs. 1 \& 2, our grating has a large $L/W$ ratio with a
period ($a$ = 70 $\mu$m) much smaller than the wavelength of the
imposed microwave ($\lambda \approx$ 1.5 cm). Similar metallic
gratings with subwavelength period have long been investigated,
which display anomalous optical effects \cite{Neviere}. Recently,
they have got renewed interest due to exotic photonic effects
showing extraordinary optical transmission \cite{Porto}. We
demonstrate here that the FM grating has its own unique
macroscopic optical behavior based on the spin dynamics. In the
simplest approximation, we treat the grating as a linear polarizer
with a permittivity tensor $\widehat{\epsilon} =
\begin{pmatrix} 1 & 0 & 0 \cr 0 & \epsilon & 0 \cr 0 & 0 &
1\end{pmatrix}$, in which we neglect the microscopic geometric
details of the patterned FM film, but focus instead on its
macroscopic characteristics of the anisotropic conductivity. By
using $\widehat{\epsilon}$ instead of $\epsilon$ in Maxwell
equations, we find that the eigenmode propagating in the FM
grating is nearly linear polarized with the wave vector given by
$k_{L}^{2}\approx\epsilon\mu_{L}k_{0}^{2}$.

The characteristics of $\mu_{L}$ plotted in Fig. 3(c) looks at
first glance similar to that of $\mu_{+}$. Indeed, both define the
same FMR since a linear polarized electromagnetic wave can be
split equally into a left and a right circular polarized wave,
with only the right one active for FMR. The characteristic
difference between $\mu_{L}$ and $\mu_{+}$ lies in the FMAR. From
$Re(\mu_{L}) = 0$, we get $\omega_{L} = \gamma \sqrt{H(H - M_0 )}$
for the FMAR, which we plot in Fig. 3(a) as the dotted curve. It
allows us to identify the photoresistance dip as the FMAR in the
FM grating. The small discrepancy left might be lifted if one
includes the details of the sample geometry \cite{geometric}.

\begin{figure} [t]
\begin{center}
\epsfig{file=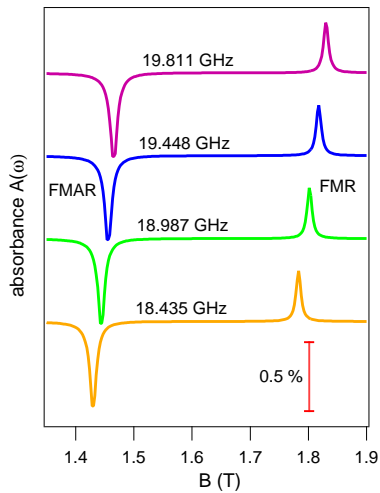,height=7 cm} \caption{(color online). The
microwave absorbance of the Py grating calculated for different
microwave frequencies. The curves are vertically offset for
clarity.} \label{Fig.4}
\end{center}
\end{figure}

Before addressing the technical and physical implication of our
results, we go a step further to calculate the absorbance
$A(\omega)$ of the Py grating on top of an insulating GaAs
substrate. The procedure is similar to that we derived recently
for a semiconductor multilayer system \cite{Bittkau}, except now
we include $\mu_{L}$ given in Eq. (2). The results of $A(\omega)$
plotted in Fig. 4 recover nicely the main feature \cite{FMAR} of
$\Delta R$ shown in Fig. 2. In particular, the agreement of the
calculated line shape for the FMR with the measured curve is
excellent, which allows us to fit accurately the dimensionless
Gilbert damping parameter $\alpha$ = 0.0075 \cite{alfa}. The
result also confirms Eq. (1), which demonstrates that the
bolometric effect in the FM metal bridges the spin dynamics and
the DC charge transport.

We summarize our work from both technical and physical point of
view. The technical difference between the photoconductivity and
transmission experiment is obvious. While a transmission
experiment measures $A(\omega)$ in Eq. (1) (or equivalently, the
\textit{high frequency surface impedance}) by monitoring the
absorption of photons, the photoconductivity experiment probes
$\Delta R$ via the change of the \textit{DC resistance} of
spin/charges. The parameter $S$ bridges both and opens free room
to enhance the sensitivity for the photoconductivity measurement.
We note that the FMAR in the FM thin film with $d < \delta$ was
unable to be detected by transmission measurements
\cite{Yager1949}. In our case where the skin depth ($\delta \sim$
1 $\mu $m) is more than one order of magnitude larger than the
thickness ($d$ = 60 nm) of the Py, the FMAR is clearly observed as
a reduction of the nonresonant photoresistance. Our technique may
also provide a new alternative means to investigate spin
excitations such as quantized spin waves, which were used to be
measured by Brillouin light scattering spectroscopy
\cite{Mathieu1998}. In principle, the photoconductivity technique
can probe the spin dissipation via $\alpha$, as well as the energy
dissipation via $\tau_{e}$, both are currently of great interest
for investigating magnetodynamics.

From the physical point of view, we uncover an intrinsic different
nature of the FMR and FMAR. While both have the common microscopic
origin of the magnetodynamic excitation with Larmor precession of
spins, FMAR is sensitive to the macroscopic geometric pattern. We
demonstrate a characteristic frequency shift of the FMAR in the
periodic FM grating from that known for FM films. Similar gratings
made of normal metals are currently of great interest for their
enhanced transmission ability based on macroscopic optical effects
\cite{Porto}. Replacing normal metals with the FM metal, one may
bring in new optical effects utilizing the strong coupling of the
electromagnetic wave with magnetodynamic excitations. The
macroscopic nature of the FMAR, together with its intrinsic nature
for enhancing transmission through FM films, could pave the way
for integrating spintronic and photonic effects using FM
microstructures.

This work is partially supported by the EU 6th-Framework Programme
through project BMR-505282, the DFG through SFB 508 and BMBF
through project 01BM905. We thank D. Heitmann, F. Giessen, D.
Grundler, H.P. Oepen, and R.E. Camley for discussions, D.
G\"{o}rlitz and G. Meier for technical help.


\begin{thebibliography}{20}

\bibitem{Tsoi2000}
M. Tsoi, \textit{et al.}, Nature {\bf 406}, 46 (2000).

\bibitem{Kiselev2003}
S. I. Kiselev, \textit{et al.}, Nature {\bf 425} 380 (2003); W.H.
Rippard, \textit{et al.}, Phys. Rev. Lett. {\bf 92}, 027201
(2004).

\bibitem{Hu2003}
C. -M. Hu, \textit{et al.}, Phys. Rev. B {\bf 67}, 201302(R)
(2003); S. Holland, \textit{et al.}, Phys. Rev. Lett. {\bf 93},
186804 (2004).

\bibitem{Neppl1979}
F. Neppl, \textit{et al.}, Phys. Rev. B {\bf 19}, 5240 (1979); K.
Hirakawa, \textit{et al.}, Phys. Rev. B {\bf 63}, 085320 (2001);
C. Zehnder, \textit{et al.}, Europhys. Lett. {\bf 63}, 576 (2003).

\bibitem{Kittel}
C. Kittel, \textit{Introduction to Solid State Physics, 6th
Edition} (John Wiley and Sons, Inc., New York, Chichester,
Brisbane, Tornonto, Singapore, 1986).

\bibitem{Yager1949}
W. A. Yager, Phys. Rev. {\bf 75}, 316 (1949); B. Heinrich and J.
F. Cochran, Phys. Rev. Lett. {\bf 29}, 1175 (1972); M. Scheffler,
M. S. thesis, University of Maryland, 1998; A. Schwartz,
\textit{et al.}, cond-mat/0010172 (2000).

\bibitem{AMR}
Due to the remanence of the superconducting coil of our magnet, a
magnetic field offset of -1.5 mT has been corrected in the plot.

\bibitem{Adeyeye1996}
A. O. Adeyeye, \textit{et al.}, J. Appl. Phys. {\bf 79}, 6120
(1996); M. Steiner, \textit{et al.}, J. Appl. Phys. {\bf 95}, 6759
(2004).

\bibitem{Hubert}
A. Hubert and R. Schaefer, \textit{Magnetic Domains - The Analysis
of Magnetic Microstructures} (Springer-Verlag, Berlin, Heidelberg,
New York, 1998).

\bibitem{Jackson}
J.D. Jackson, \textit{Classical Electrodynamics, 3rd Edition}
(John Wiley and Sons, Inc., New York, Chichester, Brisbane,
Tornonto, Singapore, 1998).

\bibitem{Nibarger2003}
J. P. Nibarger, \textit{et al.}, Appl. Phys. Lett. {\bf 83}, 93
(2003).

\bibitem{Neviere}
M. Nevi\`{e}re, in \textit{Electromagnetic Theory of Gratings},
edited by R. Petit, Springer Topics in Modern Physics  Vol. {\bf
22} (Springer-Verlag, Heidelberg, 1980).

\bibitem{Porto}
J.A. Porto, \textit{et al.}, Phys. Rev. Lett. {\bf 83}, 2845
(1999); A.P. Hibbins, \textit{et al.}, Phys. Rev. Lett. {\bf 92},
143904 (2004).

\bibitem{geometric}
To include geometric details, one needs using exact calculation as
have been nicely discussed by R.E. Camley, \textit{et al.}, Phys.
Rev. B {\bf 53}, 5481 (1996).

\bibitem{Bittkau}
K. Bittkau, \textit{et al.}, Phys. Rev. B {\bf 71}, 035337 (2005).

\bibitem{FMAR}
To reproduce the FMAR amplitude, we have phenomenologically
assumed a small energy loss ($<$ 1\%) for reflections at the
FM/GaAs interface in our calculation, which is not yet clear
whether it may reflect the influence of the surface leaky wave of
the grating as in the Wood's anomaly (Ref. [12]).
%

\bibitem{alfa}
Both the intrinsic and extrinsic contribution to $\alpha$ can be
obtained by analyzing its field dependence, which we leave to a
forthcoming longer paper.

\bibitem{Mathieu1998}
C. Mathieu, \textit{et al.}, Phys. Rev. Lett. {\bf 81}, 3968
(1998); Z.K. Wang, \textit{et al.}, Phys. Rev. Lett. {\bf 89},
027201 (2002); K. Perzlmaier, \textit{et al.}, Phys. Rev. Lett.
{\bf 94}, 057202 (2005).

\end{thebibliography}
\end{document}